\documentclass[aps,pra,reprint]{revtex4-1}

\usepackage{graphicx}
\usepackage[english]{babel}
\usepackage{amsmath, amssymb, amsfonts}
\usepackage{bm}
\usepackage[dvipsnames]{xcolor}

\newlength{\figwidth}
\newlength{\lfig}
\newlength{\sfig}
\newcommand{\RbCs}{$^{87}$Rb$^{133}$Cs}

\begin{document}
\title{Microwave shielding of ultracold polar molecules \\ with imperfectly circular polarization}
\date{\today}
\author{Tijs Karman}
\altaffiliation{Present address: ITAMP, Harvard-Smithsonian Center for Astrophysics, Cambridge, MA, 02138, USA}
\affiliation{Joint Quantum Centre (JQC) Durham-Newcastle, Department of Chemistry, Durham University, South Road, Durham, DH1 3LE, United Kingdom}
\author{Jeremy M. Hutson}
\affiliation{Joint Quantum Centre (JQC) Durham-Newcastle, Department of Chemistry, Durham University, South Road, Durham, DH1 3LE, United Kingdom}

\begin{abstract}
We investigate the use of microwave radiation to produce a repulsive shield
between pairs of ultracold polar molecules and prevent collisional losses that
occur when molecular pairs reach short range. We carry out coupled-channels
calculations on RbCs+RbCs and CaF+CaF collisions in microwave fields. We show
that effective shielding requires predominantly circular polarization, but can
still be achieved with elliptical polarization that is around 90\% circular.
\end{abstract}

\maketitle

\section{Introduction}

Ultracold polar molecules provide many new opportunities for quantum science,
and may in future provide a new platform for quantum technology. Several groups
have succeeded in producing samples of such molecules, either by direct laser
cooling of preexisting molecules \cite{Truppe:MOT:2017, Anderegg:2018} or by
associating pairs of ultracold atoms and transferring the resulting weakly
bound molecules to the ground state \cite{Ni:KRb:2008, Lang:ground:2008,
Takekoshi:RbCs:2014, Molony:RbCs:2014, Park:NaK:2015, Guo:NaRb:2016,
Rvachov:2017, Seesselberg:2018}.

A few groups have carried out experiments on collisions of ultracold polar
molecules. Fermionic molecules in identical states are protected against
collisions at the lowest temperatures \cite{Ospelkaus:react:2010,
Rvachov:2017}, but all bosonic species investigated so far have been found to
undergo fast collisional trap loss. The same is true for fermionic molecules in
different internal states. For some molecules, such as KRb
\cite{Ospelkaus:react:2010}, the loss may be attributed to energetically
allowed 2-body reactions. However, fast loss with second-order kinetics appears
to exist \cite{Ye:2018, Gregory:RbCs-collisions:2019} even when there are no
energetically allowed 2-body pathways \cite{Zuchowski:trimers:2010}. In such
cases the loss may be mediated by formation of long-lived collision complexes,
followed by either collision with a third molecule \cite{Mayle:2013} or
excitation by the trapping laser \cite{Christianen:2019b,Christianen:2019c}.

It thus appears that active measures are needed to stabilize gases of ultracold
polar molecules against collisions. An appealing possibility is microwave
shielding with blue-detuned microwave radiation \cite{Gorshkov:2008,
Karman:shielding:2018}, which provides a repulsive shield that prevents the
molecules coming close together. We have recently shown that such shielding can
be achieved with circularly polarized microwaves at high Rabi frequencies, but
not with linearly polarized microwaves \cite{Karman:shielding:2018}.

Pure circular polarization is hard to achieve for microwaves at the frequencies
and intensities needed for shielding. In this paper we explore the effect of
imperfectly circular polarization and show that good shielding can be achieved
with elliptical polarization that is around 90\% circular.

\section{Theory}
The calculations carried out in the present paper use an extension of the
methods described in Ref.~\cite{Karman:shielding:2018}. We consider a collision
between two polar molecules, each dressed with microwave radiation that is
blue-detuned by $\Delta$ from its $n=0\rightarrow1$ rotational transition. The
intensity of the microwave radiation is specified by the Rabi frequency
$\Omega$. The role of electron and nuclear spins is discussed below.

Our basic physical picture is that two molecules interact at long range via the
dipole-dipole interaction. In the presence of blue-detuned microwave radiation,
this provides a repulsive shield. Under conditions where shielding is
effective, most molecules are reflected at long range, and do not approach each
other close enough for short-range processes to occur. There are nevertheless
two processes that can cause loss. First, any colliding pairs that do reach
short range are likely to be lost. We characterize this by the probability of
reaching short range (RSR), which may be expressed as a rate coefficient only
in the limit of 100\% loss at short range. Secondly, even pairs that are
reflected at long range may undergo inelastic transitions to lower-lying
field-dressed states. These transfer internal energy into relative kinetic
energy and generally eject both collision partners from the trap. We refer to
this latter process as microwave-induced loss.

To model shielding and loss, we carry out coupled-channels scattering
calculations. We propagate two sets of linearly independent solutions of the
coupled-channels equations, using the renormalized Numerov method
\cite{Johnson:1978}, and apply both capture boundary conditions at short range
and $S$-matrix boundary conditions at long range \cite{light:76, clary:87,
janssen:13}. We calculate both the probability of reaching short range (RSR)
and the rate coefficient for microwave-induced loss. The remaining flux is
shielded and scatters elastically.

The molecules are modeled as rigid rotors with a dipole moment. The monomer
Hamiltonian of molecule $X$ is
\begin{align}
\hat{H}^{(X)} = b_\mathrm{rot} \hat{n}^2 + \hat{H}_\mathrm{hyperfine}^{(X)}
+ \hat{H}_\mathrm{ac}^{(X)}.
\label{eq:monH}
\end{align}
The first term describes the rotational kinetic energy, with rotational
constant $b_\mathrm{rot}$. The second term describes the molecular hyperfine
Hamiltonian, and is discussed in the Supplement of
Ref.~\cite{Karman:shielding:2018}. The last term represents the interaction
with a microwave electric field \cite{Cohen-Tannoudji:API:1998},
\begin{align}
\hat{H}_\mathrm{ac}^{(X)} = -\sqrt{\frac{\hbar\omega}{2\epsilon_0 V_0}}
\left[ \hat{\mu}_\sigma^{(X)} \hat{a}_\sigma +
\hat{\mu}_\sigma^{(X)\dagger} \hat{a}_\sigma^\dagger\right]
	+ \hbar\omega  \left(\hat{a}_\sigma^\dagger \hat{a}_\sigma - N_0\right).
\label{eq:Hac}
\end{align}
Here, $N_0 = \epsilon_0 E_\mathrm{ac}^2 V_0 /2\hbar\omega$ is the reference
number of photons in a reference volume, $V_0$, at microwave electric field
strength, $E_\mathrm{ac}$ \cite{Avdeenkov}.
The operators
$\hat{a}_\sigma^\dagger$ and $\hat{a}_\sigma$ are creation and annihilation
operators for photons in polarization mode $\sigma$ and angular frequency
$\omega$. The polarization may be linear, with Cartesian components $\sigma =
x$, $y$, and/or $z$, circular, with spherical components $\sigma = \pm1$ and
$\hat{\mu}^{(X)}_{\pm 1} = \mp \left( \hat{\mu}^{(X)}_x \pm i
\hat{\mu}^{(X)}_y\right)/\sqrt{2}$, or elliptical. A general elliptical
polarization in the $xy$ plane can be described as
\begin{align}
	\sigma(\xi) = \sigma_+ \cos\xi - \sigma_- \sin\xi.
\end{align}
The ellipticity angle $\xi$ interpolates between pure right-hand circular
polarization, $\sigma_+$, at $\xi=0$ and linear x polarization, $\sigma_x$, at
$\xi=\pi/4$. The form of the interaction with a microwave electric field,
Eq.~\eqref{eq:Hac}, remains valid with the substitution $\hat{\mu}_{\sigma}
\rightarrow \hat{\mu}_{+1} \cos\xi - \hat{\mu}_{-1} \sin\xi$.

The total Hamiltonian is
\begin{align}
\hat{H} = -\frac{\hbar^2}{2M} \frac{1}{R} \frac{d^2}{dR^2} R
+ \frac{\hbar^2 \hat{L}^2}{2M R^2} + \hat{H}^{(A)} + \hat{H}^{(B)} + \hat{V}(R).
\label{eq:Htot}
\end{align}
Here $M$ is the reduced mass, $R$ is the distance between the molecules and
$\hat{L}$ is the dimensionless angular momentum operator associated with the
end-over-end rotation of the intermolecular axis, $\vec{R}$. The first term
describes the radial kinetic energy and the second the centrifugal kinetic
energy. The third and fourth terms correspond to the monomer Hamiltonian of
Eq.\ (\ref{eq:Hac}). The final term is the interaction potential $\hat{V}(R)$,
which in the present work is limited to the dipole-dipole interaction
\cite{Karman:shielding:2018},
\begin{align}
\hat{V}(R) = -\frac{\sqrt{6}}{4\pi\epsilon_0 R^3}
T^{(2)}(\hat{\mu}^{(A)},\hat{\mu}^{(B)}) \cdot C^{(2)}(\hat{R}),
\end{align}
where $C^{(2)}(\hat{R})$ and $T^{(2)}(\hat{\mu}^{(A)},\hat{\mu}^{(B)})$ are
second-rank spherical tensors whose normalisations and scalar product are given
in the Supplemental Material of ref.\ \cite{Karman:shielding:2018}.

We use completely uncoupled basis sets. The basis set for monomer $X=A,B$
consists of products of rotational, spin and photon states,
\begin{align}
|n_X m_{n_X}\rangle |i_X m_{i_X}\rangle |N\rangle,
\end{align}
where $|i_X m_{i_X}\rangle$ schematically represents all the spins
on molecule $X$; it may be a product of spin functions, including electron spin
where necessary.
The blue detuning is given by positive $\Delta$.
For the colliding pair of molecules we introduce an angular momentum state, $|L
M_L\rangle$, that describes the end-over-end rotation of the intermolecular
axis. The basis functions for the pair of molecules are
\cite{hanna:10,owens:16, owens:17}
\begin{align}
|n_A m_{n_A}\rangle |i_A m_{i_A}\rangle |n_B m_{n_B}\rangle |i_B m_{i_B}\rangle |L M_L\rangle|N\rangle.
\end{align}
Only even values of $L$ are included: The only interaction that couples states
with different $L$ is the dipole-dipole interaction, and this conserves the
parity of $L$. The basis functions are adapted to permutation symmetry as
described in the Supplement of Ref.\ \cite{Karman:shielding:2018}.

Excluding the photons, the direct-product basis functions have a well-defined
projection quantum number $M_\mathrm{tot}$ for the projection of the total
angular momentum along the space-fixed $z$ axis,
\begin{align}
M_\mathrm{tot} = m_{n_A} + m_{i_A} + m_{n_B} + m_{i_B} + M_L.
\end{align}
For linear $z$ or circular $x\pm i y$ polarizations, with $\sigma=0$, $\pm1$,
the generalized projection quantum number $\mathcal{M}$ is conserved,
\begin{align}
\mathcal{M} = M_\mathrm{tot} + \sigma N,
\label{eq:mathcalM}
\end{align}
and the basis set can be limited to include only functions with a single value
of $\mathcal{M}$. For elliptical polarization in the $xy$ plane, however, there
is no generalized projection quantum number that is conserved. In this case,
the basis set must include functions with various values of $\mathcal{M}$, and
the calculation is correspondingly more computer intensive.

We showed in Ref.~\cite{Karman:shielding:2018} that molecular fine and
hyperfine interactions cause an increase in microwave-induced loss at zero
magnetic field, but that the increase can be suppressed by applying a moderate
magnetic field perpendicular to the plane of polarization. We demonstrate in
the Supplemental Material that the field required to suppress the effects of
hyperfine interactions is proportional to the ratio of the hyperfine constants
to the nuclear g-factors. For molecules in $^2\Sigma$ states, the field needed
is proportional to the ratio of the spin-rotation constant to the electron
g-factor. Since there is a tendency for both hyperfine coupling constants and
nuclear g-factors to increase with atomic number, the magnetic field required
to suppress hyperfine effects does not vary enormously, and is typically on the
order of 100~G. The key issue is that the magnetic field must be high enough
that $m_n$ is nearly conserved. When it is, the microwave-induced loss for the
bialkalis is reduced to the hyperfine-free level, while that for molecules in
$^2\Sigma$ states is incompletely suppressed \cite{supplement}.

\section{Results \label{sec:results}}

\begin{figure}[t]
\begin{center}
\includegraphics[width=\figwidth,clip]{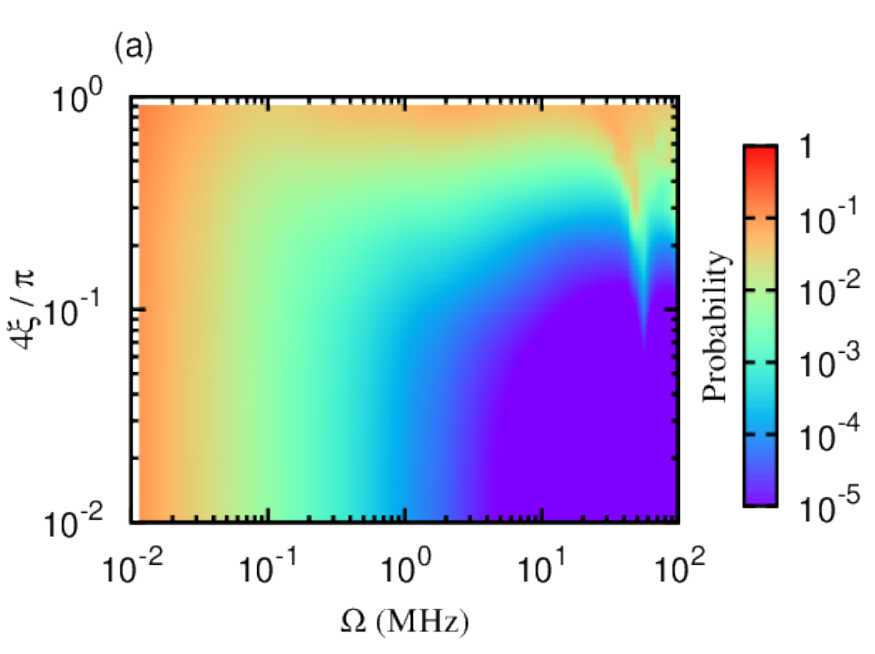}
\includegraphics[width=\figwidth,clip]{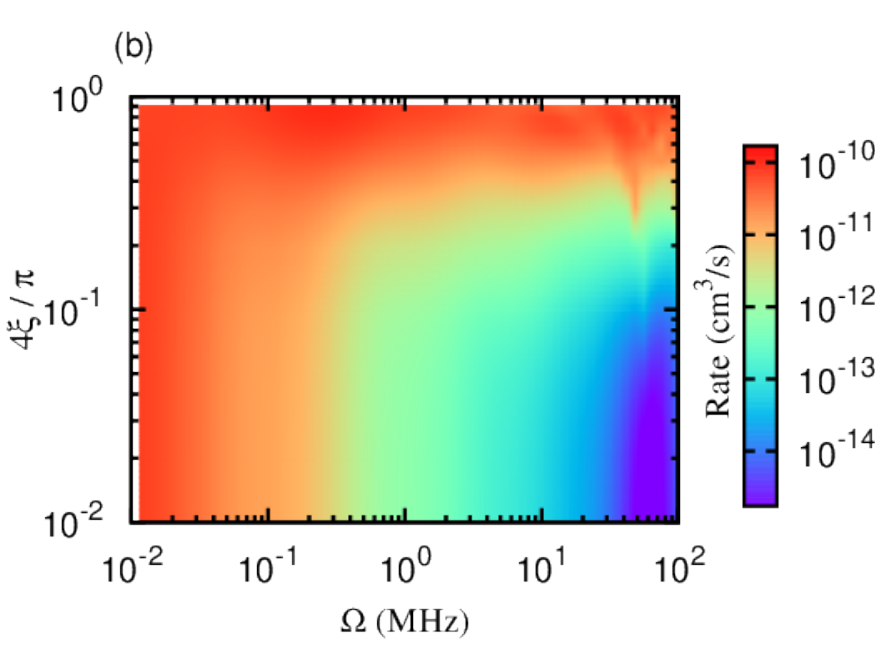}
\caption{ \label{fig:elliptical_RbCs}
Probability of RSR (a) and rate coefficient for microwave-induced loss (b) in
RbCs+RbCs collisions with microwave detuning $\Delta=0$, as a function of Rabi
frequency, $\Omega$, and ellipticity angle, $\xi$. }
\end{center}
\end{figure}

We have carried out coupled-channel calculations for RbCs+RbCs and CaF+CaF
collisions in microwave fields, with the molecules in their ground
electronic states, $X^1\Sigma^+$ and $X^2\Sigma^+$, respectively. The
calculations used $n_\textrm{max}=1$, $L_\textrm{max}=6$, $N=0,-1,-2$, and a
collision energy of 1~$\mu$K. Electron and nuclear spins were not
included explicitly, but for CaF the calculation included spin-dependent
couplings averaged over the spin-stretched state. As shown in the Supplemental
Material, these calculations are appropriate for magnetic fields above 100~G
\cite{supplement}.

Figure~\ref{fig:elliptical_RbCs} shows loss processes for RbCs as a function of
the ellipticity angle, $\xi$, and the Rabi frequency, $\Omega$, for fixed
$\Delta=0$. Panel (a) shows the probability of RSR, and panel (b) shows the
rate coefficient for microwave-induced loss, i.e., inelastic transitions to
lower-lying field-dressed states. Figure~\ref{fig:elliptical_RbCs} shows the
probability of RSR and the rate coefficient for microwave-induced loss for RbCs
as a function of the ellipticity angle, $\xi$, and the Rabi frequency,
$\Omega$, for fixed $\Delta=0$. At the top of the figures, where
$4\xi/\pi\approx 1$, the probability of RSR and the microwave-induced loss rate
are both large. This corresponds to linear polarization, for which shielding is
ineffective. At the bottom of the figure, below $4\xi/\pi\approx 0.01$, the
probability of RSR and the microwave-induced loss rate become independent of
$\xi$ and visually indistinguishable from the result for circular polarization.
Losses can be suppressed to four orders of magnitude below the universal loss
rate \cite{Idziaszek:PRL:2010}, which for RbCs at zero energy is $1.7\times
10^{-10}$~cm$^3~$s$^{-1}$. These values of $\xi$ may be experimentally
realizable. An ellipticity of $4\xi/\pi\approx 0.1$, corresponding a microwave
field that is 90 \% circular and a power ratio around 20~dB between $\sigma_+$
and $\sigma_-$ polarizations, results in somewhat increased losses but still
provides effective shielding.

\begin{figure}[b]
\begin{center}
\includegraphics[width=\figwidth,clip]{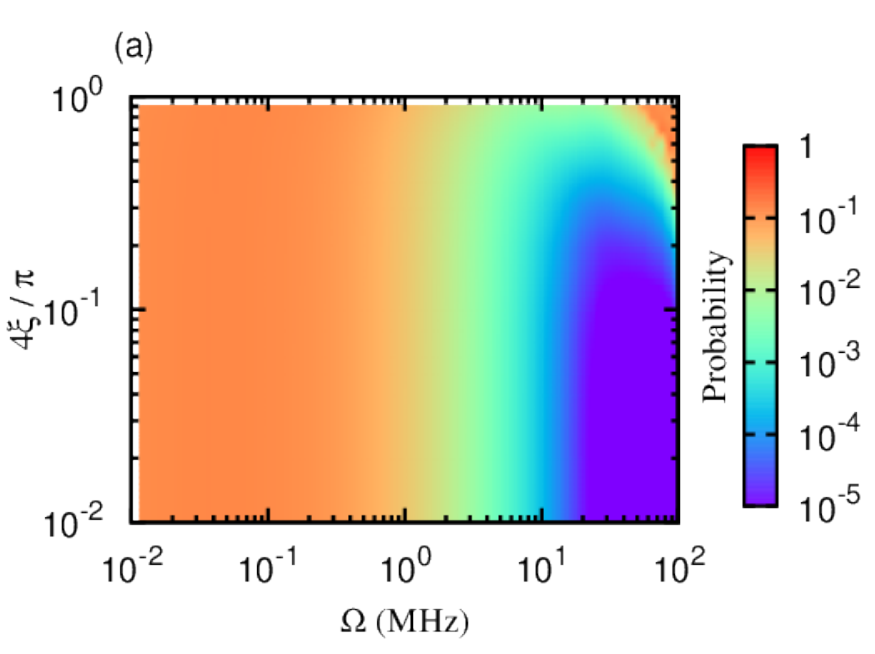}
\includegraphics[width=\figwidth,clip]{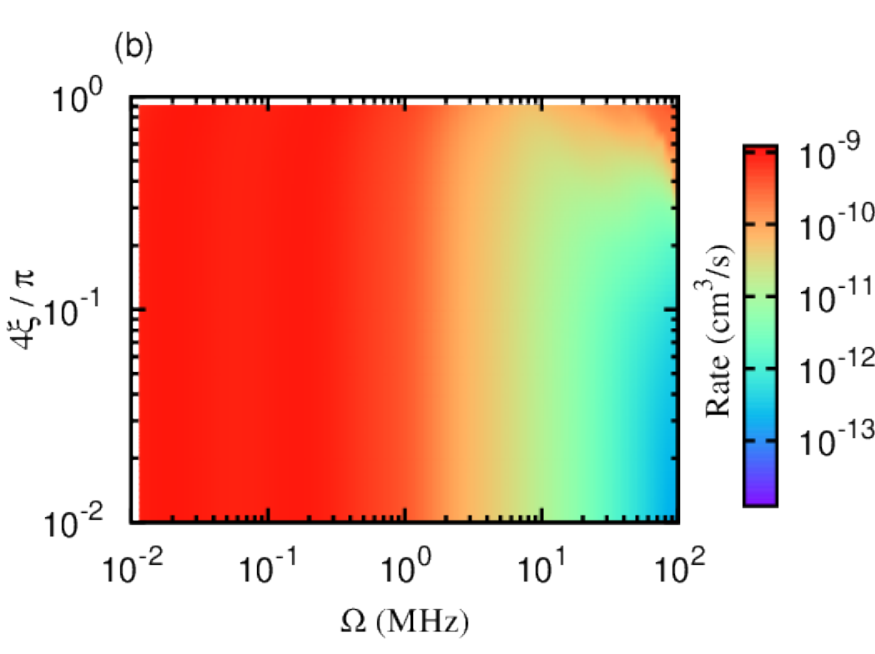}
\caption{ \label{fig:elliptical_CaF}
Probability of RSR (a) and rate coefficient for microwave-induced loss (b) in
CaF+CaF collisions with microwave detuning $\Delta=0$, as a function of Rabi
frequency, $\Omega$, and ellipticity angle, $\xi$. }
\end{center}
\end{figure}

Figure~\ref{fig:elliptical_CaF} shows the probability of RSR and the rate
coefficient for microwave-induced rate for CaF as a function of the ellipticity
angle, $\xi$, and the Rabi frequency, $\Omega$, for fixed $\Delta=0$. The spin
degrees of freedom are not explicitly accounted for, but the calculation
includes fine and hyperfine couplings averaged over the spin-stretched state,
which is appropriate for magnetic fields above 100~G \cite{supplement}. The
resulting couplings increase the microwave-induced loss. The resulting
shielding is less effective than for RbCs, but the losses are still reduced by
up to three orders of magnitude from the universal loss rate of $5\times
10^{-10}$~cm$^3$~s$^{-1}$. The loss rates are essentially indistinguishable
from those for circular polarization for $4\xi/\pi \le 0.1$.

\section{Fixed-$\theta$ collisions (sudden approximation)}

To understand the way that the losses depend on the ellipticity angle, we
perform additional calculations using a sudden approximation in which the
orientation of the intermolecular axis is assumed to remain constant during a
collision. The orientation is specified by the polar angles ($\theta$, $\phi$)
of $\hat{R}$ with respect to the space-fixed frame. The spherical harmonics
$|L, M_L \rangle$ are dropped from the basis set. This approximation is
expected to be accurate if the period of an end-over-end rotation of the
complex is long compared to the duration of a collision, which may not be the
case. It also neglects nonadiabatic losses due to couplings involving
$d/d\theta$, which may be significant. The fixed-$\theta$ approximation is
therefore not quantitatively accurate, but it nevertheless provides useful
insights. The results presented in Sec.~\ref{sec:results} were obtained from
full coupled-channels calculations that do not use the sudden approximation.

\begin{figure*}
\begin{center}
\includegraphics[width=0.475\textwidth,clip]{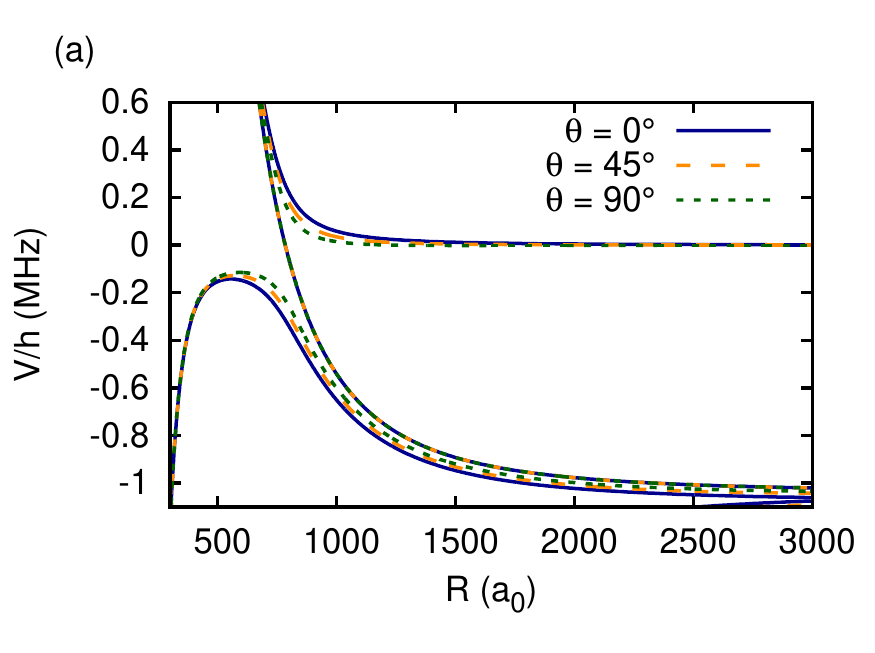}
\includegraphics[width=0.475\textwidth,clip]{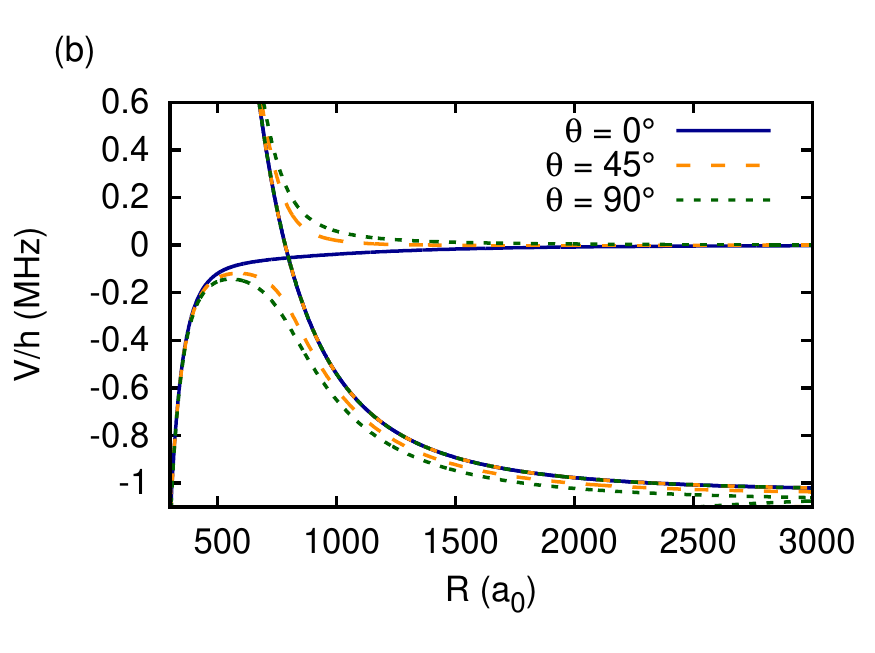}
\includegraphics[width=0.475\textwidth,clip]{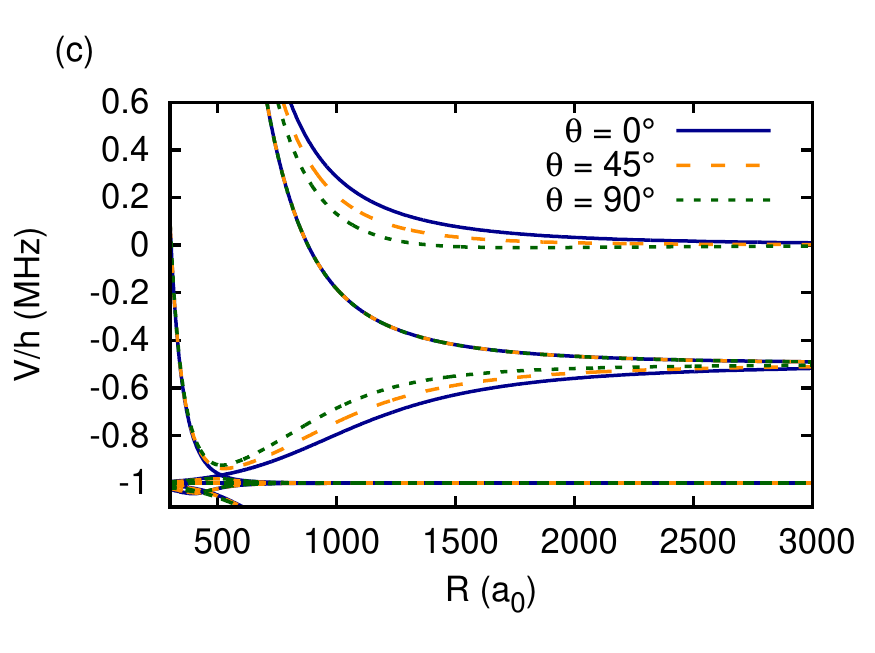}
\includegraphics[width=0.475\textwidth,clip]{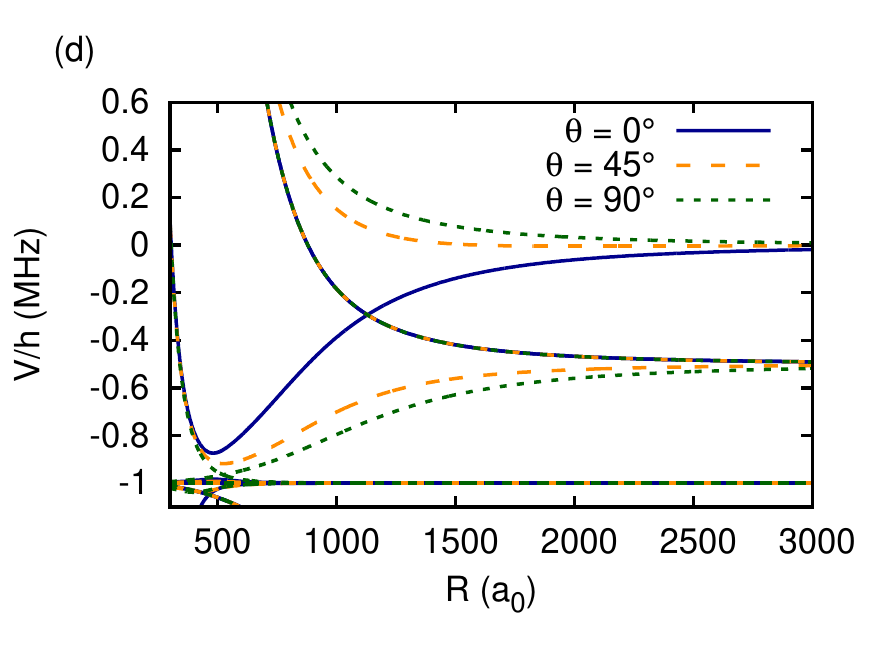}
\caption{ \label{fig:fixedq_pot} Adiabatic potential curves for RbCs+RbCs with
$\Omega=0.2$~MHz and fixed orientation $\theta$ as indicated in the legend. The
different panels correspond to: (a) Circular polarization for
$\Omega\ll\Delta$, (b) linear polarization for $\Omega\ll\Delta$, (c) circular
polarization for $\Omega\gg\Delta$, and (d) linear polarization for
$\Omega\gg\Delta$. }
\end{center}
\end{figure*}

Figure~\ref{fig:fixedq_pot} shows adiabatic potential curves for RbCs+RbCs,
defined as eigenvalues of the pair Hamiltonian, Eq.~\eqref{eq:Htot}, excluding
radial kinetic energy, as a function of $R$ for fixed $\theta$. This
representation is much simpler than is obtained by diagonalizing in the full
basis set that includes the dependence on $\theta$ through the partial-wave
expansion, as shown in Fig.~\ref{fig:fixedq_pot} of the Supplement of
Ref.~\cite{Karman:shielding:2018}. Figure~\ref{fig:fixedq_pot} provides a less
crowded representation of the adiabats that removes many of the inconsequential
crossings of low and high-$L$ states corresponding to different thresholds.
In each case the colliding molecules are initially at the
uppermost of the thresholds shown.

The upper two panels of Figure~\ref{fig:fixedq_pot} show adiabats for the
off-resonant case, $\Omega\ll\Delta$. This is not the parameter regime in which
effective shielding is realized, but the potential curves are more easily
understood. In this case the ground-state ($|n_A,n_B,N\rangle = |0,0,0\rangle$)
and microwave-dressed excited-state ($|0,1,-1\rangle$) potentials are
essentially unmodified by the microwaves. The ground-state (upper) potential is
determined by rotational dispersion and varies with $R^{-6}$. The excited-state
(lower) potential is split by resonant dipole-dipole interactions into an
attractive and repulsive branch, which vary with $R^{-3}$; the attractive
branch is mostly off scale in Fig.\ \ref{fig:fixedq_pot}. Resonant
dipole-dipole interactions quantize the rotational angular momentum along the
intermolecular axis with projection quantum number $K$, giving potentials that
are independent of orientation with respect to the microwave polarization:
$K=0$ for the attractive branch and $K=\pm 1$ for the repulsive branch. In this
off-resonance scenario, coupling by the microwaves is significant only near the
point where the ground-state potential crosses the repulsive $K=\pm 1$ branch
of the excited state, around 800~$a_0$ in Fig.\ \ref{fig:fixedq_pot}. For
linear $\pi$ or circular $\sigma_\pm$ polarization, the microwave field couples
the ground state to the excited state with well-defined space-fixed projection
quantum number $M_N=m_{nA}+m_{nB}=0$ or $\pm 1$, respectively. Hence, the
coupling to the repulsive $K=\pm 1$ branch depends on the polar angle of the
intermolecular axis, $\theta$, as the Wigner $d$-function
$d^{(1)}_{M_N,K}(\theta)$. For circular $\sigma_\pm$ polarization, shown in
Fig.~\ref{fig:fixedq_pot}(a), the crossing is avoided for all angles $\theta$,
although the degree of avoidedness is anisotropic and varies with
$\sqrt{1+\cos^2\theta}$. For linear $\pi$ polarization, shown in
Fig.~\ref{fig:fixedq_pot}(b), the avoidedness of the crossing varies with
$\sin\theta$, and vanishes at $\theta=0$. This implies that there is a ``hole"
in the shielding at $\theta=0$ with linear polarization.

Next, we consider the resonant case, $\Omega\gg\Delta=0$, where effective
shielding is achieved with near-circular polarization. The adiabats are shown
in the lower two panels of Figure~\ref{fig:fixedq_pot}. Here, there is strong
dressing that mixes the $n=0$ ground and $n=1$ excited states even for the
isolated molecules. This can be interpreted as inducing an oscillating dipole
moment in the lab frame, which produces first-order dipole-dipole interactions
upon time averaging. As a result, the adiabats vary with $R^{-3}
P_2(\cos\theta)$ at long range. The space-fixed dipole-dipole interactions are
attractive for some orientations, i.e., near $\theta=0$ for linear $\pi$
polarization and near $\theta=90^\circ$ for circular $\sigma_\pm$ polarization.
However, at shorter separation, the dipole-dipole interaction becomes stronger
than $\hbar\Omega$, and this interaction again quantizes the molecules along
the intermolecular axis, such that the top adiabatic curve is repulsive for all
$\theta$. As discussed above, the coupling from the initial state to this
repulsive branch depends on $\theta$. For circular polarization, shown in
Fig.~\ref{fig:fixedq_pot}(c), the angle dependence is again such that the
coupling to the repulsive branch never vanishes. For linear polarization, shown
in Fig.~\ref{fig:fixedq_pot}(d), the shielding again vanishes at $\theta=0$,
where both $K=\pm1$ repulsive states are uncoupled from the initial state. For
small $\theta$, this leads to a narrowly avoided crossing between the initial
state and the lower field-dressed level; this occurs outside the repulsive
shield and is classically accessible. As a result, linear polarization does not
achieve shielding, but it does give rise to additional microwave-induced
losses. For imperfectly circular polarization, the effective coupling to the
repulsive branch is reduced from that for circular polarization, but does not
vanish.

It is worth noting that the space-fixed dipole-dipole interaction discussed
here produces attractive interactions outside the repulsive shield. Where these
support bound states, they can be used to tune the scattering length while
shielding from short-range losses, as shown in Ref.~\cite{Lassabliere:2018}.

We next carry out coupled-channels calculation of shielding and loss in the
fixed-$\theta$ approximation. For these we consider elliptical polarization in
the $xy$ plane, with $\xi=\pi/4$ corresponding to linear polarization along
$x$ ($\sigma_x$ polarization). This differs from the discussion above, which
considered linear polarization along $z$ ($\pi$ polarization). Since only the
microwave polarization defines a preferential direction, this choice is
inconsequential, but it does mean that collisions now occur along
the axis of linear polarization when ${\theta = 90^\circ}$ rather than
$\theta=0$.

\begin{figure*}
\begin{center}
\includegraphics[width=0.4750\textwidth,clip]{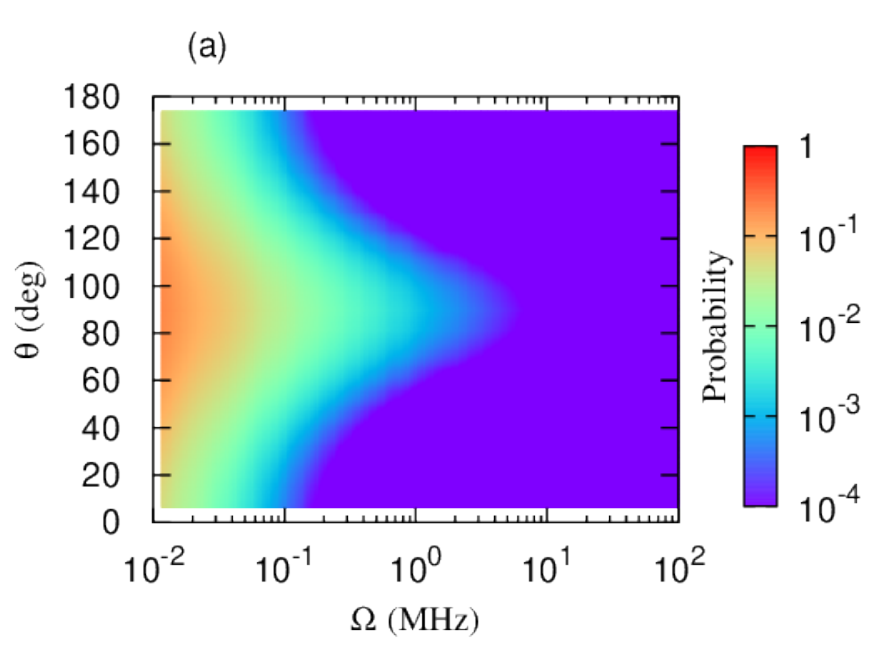}
\includegraphics[width=0.4750\textwidth,clip]{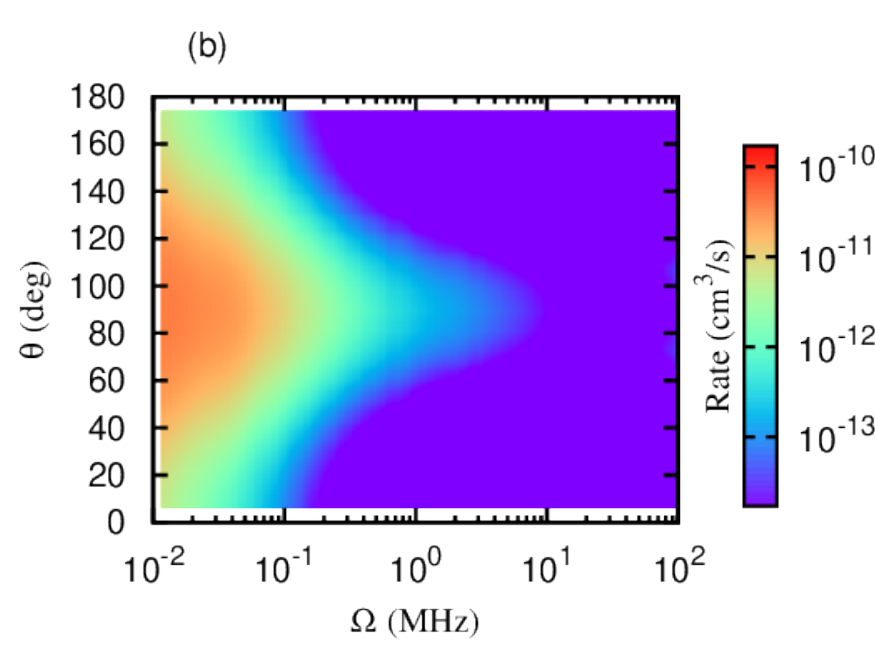}
\includegraphics[width=0.4750\textwidth,clip]{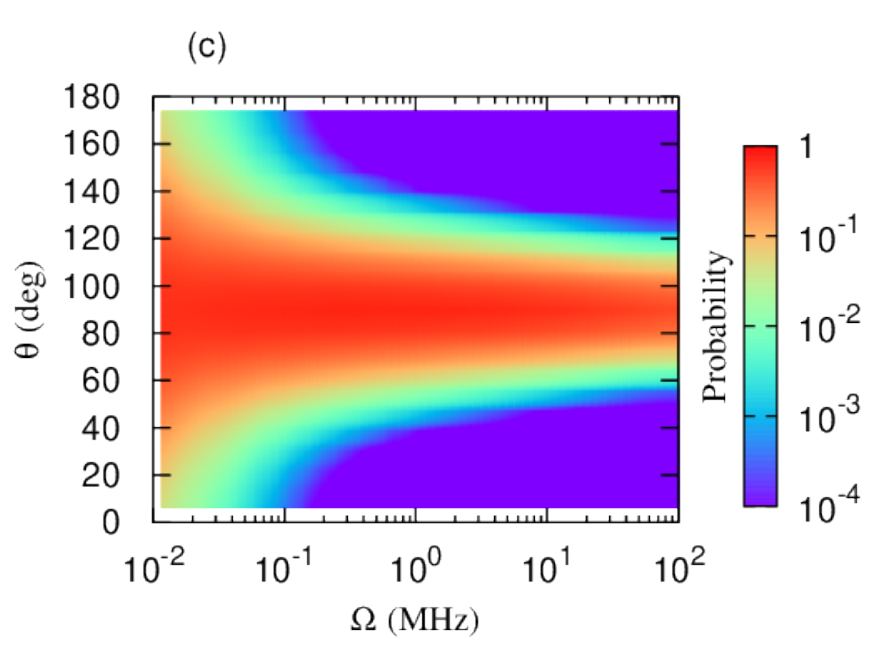}
\includegraphics[width=0.4750\textwidth,clip]{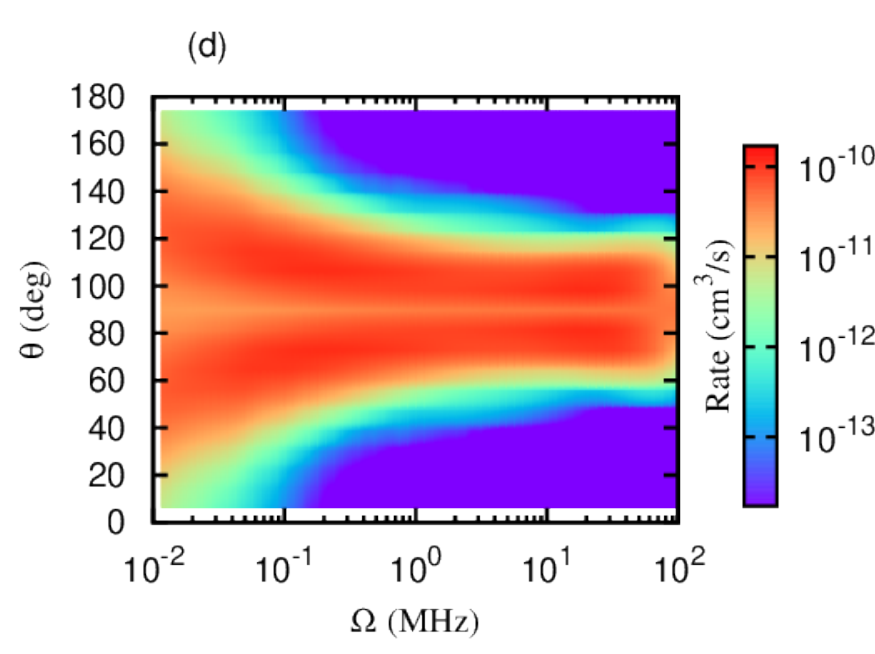}
\includegraphics[width=0.4750\textwidth,clip]{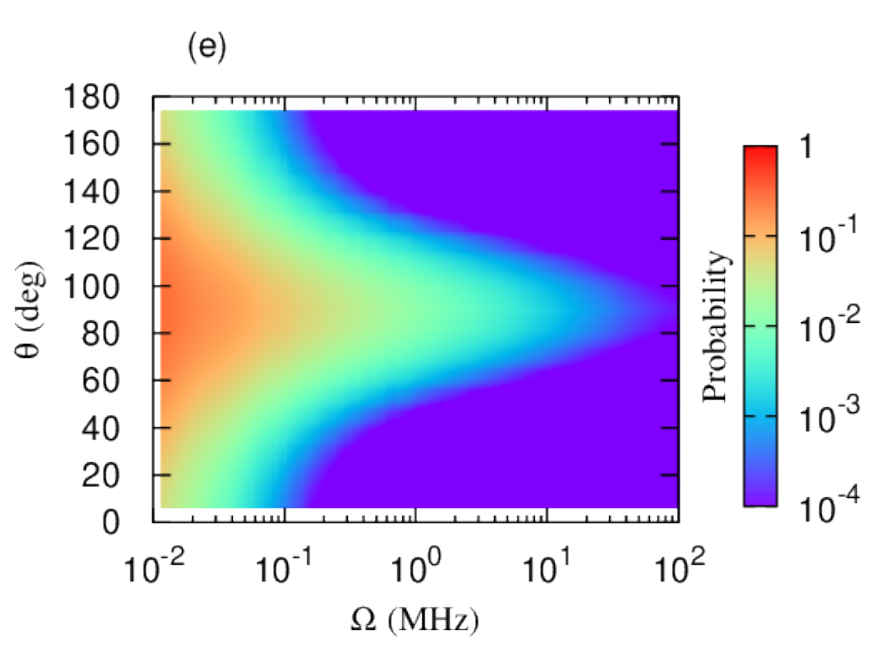}
\includegraphics[width=0.4750\textwidth,clip]{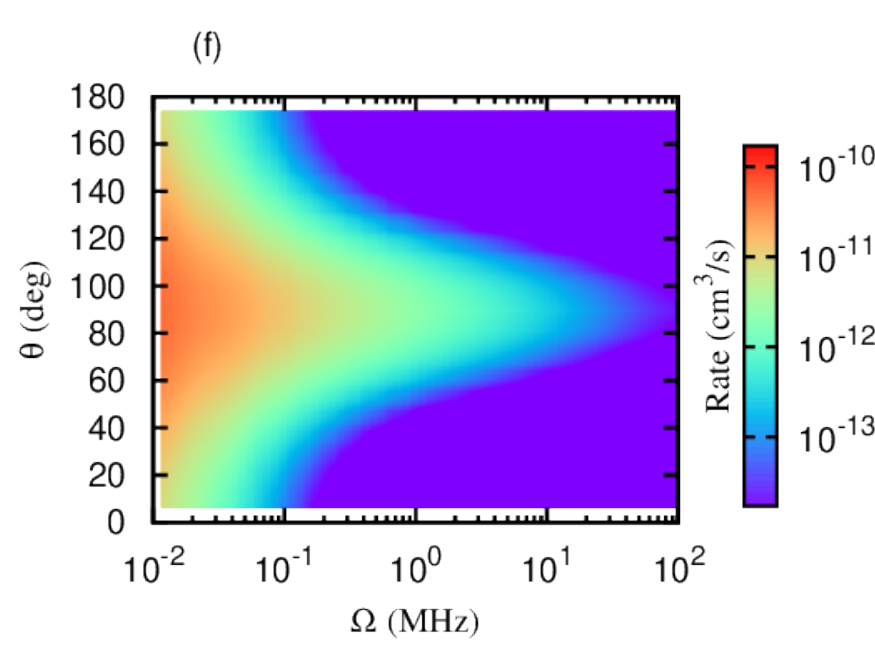}
\caption{ \label{fig:fixedqloss} Probabilities for reaching short range
[left-hand column, panels (a,c,e)] and rates of microwave-induced loss
[right-hand column, panels (b,d,f)] as a function of Rabi frequency and fixed
orientation $\theta$. Panels (a,b) are for $\sigma^+$ circular polarization,
(c,d) are for $\sigma_x$ linear polarization, and panels (e,f) are for
elliptical polarization with $\xi= 0.2 \pi/4$.}
\end{center}
\end{figure*}

Figure~\ref{fig:fixedqloss} shows the probabilities of RSR and the rate
coefficients for microwave-induced loss in RbCs+RbCs collisions, for
$\Delta=0$, as a function of the Rabi frequency, $\Omega$, and the fixed angle,
$\theta$, between the intermolecular axis and microwave propagation direction.
The three rows correspond to different polarizations: $\sigma_+$, $\sigma_x$,
and elliptical polarization with $\xi= 0.2 \pi/4$. For circular polarization,
shown in the top row, we find both types of loss can be suppressed by using
sufficiently high Rabi frequency. The loss is highest for $\theta=90^\circ$, where
coupling to the repulsive branch of the resonant dipole-dipole interaction is
weakest and long-range space-fixed dipole-dipole interactions are attractive.
For linear $x$ polarization, shown in the center row, we find losses are also
highest near $\theta=90^\circ$, where unshielded collisions along the polarization
direction occur, as discussed above. These losses cannot be reduced by
increasing the Rabi frequency. The bottom row shows results for elliptical
polarization with $\xi= 0.2 \pi/4$. The results are qualitatively similar to
those in the case of circular polarization, but somewhat higher Rabi
frequencies are required to suppress the loss.

\section{Conclusions}

We have investigated shielding of ultracold molecular collisions using
microwave radiation with imperfectly circular polarization. The goal of
microwave shielding is to prevent colliding pairs of molecules from reaching
short range, where losses occur with high probability, while simultaneously
avoiding long-range losses induced by the microwaves themselves. We have
carried out coupled-channels calculations on RbCs+RbCs and CaF+CaF collisions
to evaluate both the probabilities of reaching short range and the rate
coefficients for microwave-induced loss.

We showed previously \cite{Karman:shielding:2018} that effective shielding can
be achieved with circularly polarized microwaves but not with linearly
polarized microwaves. However, pure circular polarization is hard to achieve in
an apparatus designed for trapping ultracold molecules. Here we investigate how
the effectiveness of shielding degrades for imperfectly circular polarization.
We show that effective shielding can still be achieved with elliptical
polarization corresponding to a microwave field that is around 90\% circular,
with a power ratio around 20~dB between $\sigma_+$ and $\sigma_-$
polarizations.

Molecular fine and hyperfine structure can interfere with microwave shielding.
We show that, for molecules in $^1\Sigma$ states, such as the bialkalis,
effective shielding can be restored by applying a moderate magnetic field, of
order 100~G, perpendicular to the plane of polarization. For molecules in
$^2\Sigma$ states, such as CaF, shielding is not fully restored by a magnetic
field but losses can still be reduced by up to three orders of magnitude
compared to the universal limit of complete short-range loss.

We interpret our results in terms of an approximate model in which collisions
occur at a fixed orientation of the intermolecular axis. This simplifies the
effective potential curves that govern shielding and loss. For linear
polarization, there are two curves in this model that cross (without an avoided
crossing) when a collision occurs with the intermolecular axis along the axis
of polarization. This is responsible for the lack of shielding for linear
polarization. Elliptical polarization turns this crossing into an avoided
crossing, which increases in strength as the degree of circular polarization
increases. This restores effective shielding even for imperfectly circular
polarization.

We anticipate that microwave shielding with near-circular polarization will be
a valuable tool for reducing collisional losses in samples of ultracold polar
molecules.

\section{Acknowledgement}
This work was supported by the U.K. Engineering and Physical Sciences Research
Council (EPSRC) Grants No.\ EP/P008275/1, EP/N007085/1 and EP/P01058X/1.

\nocite{data}
\bibliography{../../bibfile,../../../all,../bibfile}

\renewcommand{\thefigure}{S\arabic{figure}}
\setcounter{figure}{0}
\clearpage
\onecolumngrid
\vspace{\columnsep}

        \begin{center}
        \large{\textbf{Supplementary Information for: \\ Microwave shielding of ultracold polar molecules \\ with imperfectly circular polarization}}

        Tijs Karman$^{\ast}$, Jeremy M. Hutson

        \textit{Joint Quantum Centre (JQC) Durham-Newcastle, Department of Chemistry,\\ Durham University, South Road, Durham, DH1 3LE, United Kingdom}

        \end{center}

\vspace{\columnsep}
\twocolumngrid

\section{Suppression of the effects of fine and hyperfine structure by magnetic
field}

\begin{figure}
\begin{center}
\includegraphics[width=\columnwidth,clip]{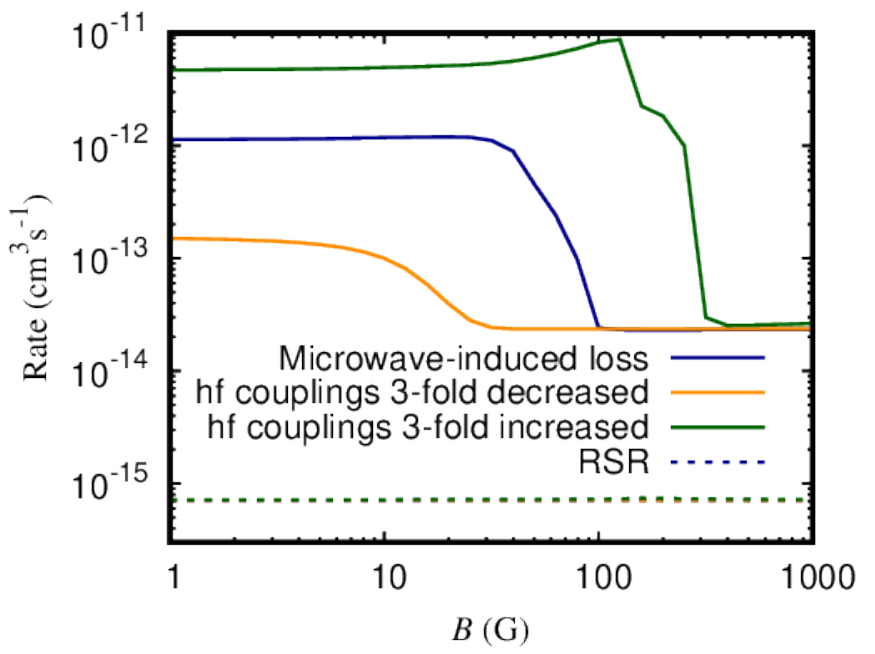}
\caption{ \label{fig:hyperfine2} Rate coefficients for RSR and
microwave-induced loss in RbCs+RbCs collisions, as a function of the magnetic
field strength, $B$. The loss rates shown are obtained using the physical
hyperfine constants, as well as hyperfine constants scaled up or down by a
factor 3. For scaled hyperfine interactions, the suppression occurs at
approximately $B=30$~G and $300$~G, rather than 100~G. The effect of hyperfine
interactions on the microwave-induced loss rate also scales with the magnitude
of the hyperfine coupling constants.}
\end{center}
\end{figure}

Figure~\ref{fig:hyperfine2} shows rate coefficients for RSR (assuming 100\%
loss at short range) and microwave-induced loss for RbCs+RbCs collisions,
including hyperfine structure, as a function of magnetic field perpendicular to
the plane of circularly polarized microwave radiation. Results are shown for
physical \RbCs, as well as for RbCs with hyperfine interactions scaled up and
down by a factor of three, respectively. The magnetic fields required to
suppress the effect of hyperfine interactions are found also to increase and
decrease by a factor three, respectively. The Zeeman interactions required to
quantize the nuclear spins along the magnetic field are thus proportional to
the magnetic field. This explains why suppression of the effects of fine and
hyperfine structure was found to occur at approximately the same magnetic
field, $B\approx 100$~G, for RbCs, KCs, and CaF. The hyperfine interactions are
weaker for KCs than for RbCs, whereas for CaF the fine and hyperfine
interactions are orders of magnitude stronger. However, the nuclear and
electronic $g$-factors scale by approximately the same factors. Hence, the
Zeeman interaction is weaker (stronger) for KCs (CaF) by approximately the same
factor, such that similar magnetic fields provide suppression of the effects of
hyperfine interactions in the three systems.

In the low-field limit, the enhancement of microwave-induced loss due to fine
and hyperfine interactions also increases with the strength of the hyperfine
interactions, as can be seen in Fig.~\ref{fig:hyperfine2}. This is also in line
with Ref.~\cite{Karman:shielding:2018}, where it was found that the enhancement
of microwave-induced loss is smaller for KCs than for RbCs, and much larger for
CaF than for RbCs.

\begin{figure}
\begin{center}
\includegraphics[width=\columnwidth,clip]{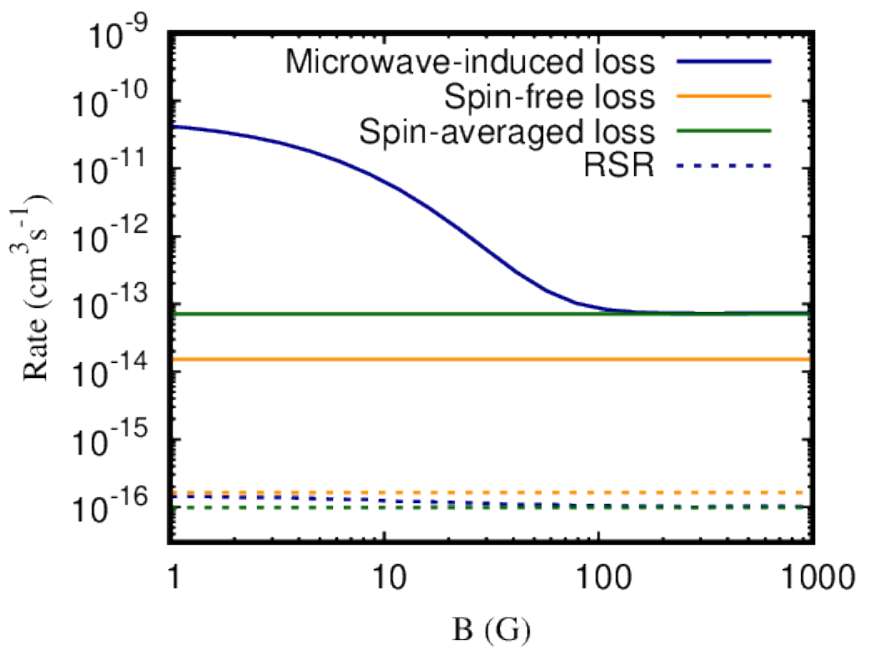}
\caption{ \label{fig:CaF_highB} Rate coefficients for RSR and microwave-induced
loss in CaF+CaF collisions, as a function of the magnetic field strength, $B$.
Loss rates shown are obtained from three sets of calculations: A full
calculation that includes the fine and hyperfine structure, a spin-free
calculation that neglects fine and hyperfine interactions altogether, and a
spin-averaged calculation in which fine and hyperfine interactions are averaged
over the spin-stretched initial hyperfine state. }
\end{center}
\end{figure}

In the high-field limit, the fine and hyperfine enhancement of
microwave-induced loss is suppressed because the spins quantize along the
magnetic field. The spin projections effectively become spectator quantum
numbers, and only the initial spin state contributes. Fine and hyperfine terms
averaged over this spin state do exist, although they are not present in the
spin-free calculation. For example, for a spin $|S M_S\rangle$ along the
magnetic field axis, the spin-rotation interaction, $\gamma \hat{N} \cdot
\hat{S}$, is averaged to a term ${\langle S M_S|\gamma \hat{N} \cdot \hat{S} |
S M_S \rangle =  \gamma M_S \hat{N}_z}$, which does not occur in the spin-free
case. For the bialkalis RbCs and KCs, $\gamma \ll \Omega$. The averaged
hyperfine terms can thus be neglected and the loss rates are suppressed to the
hyperfine-free level. For CaF, however, the fine and hyperfine interactions are
not negligible compared to $\Omega=100$~MHz; for example, the
electron-spin-rotation constant is $\gamma\approx 40$~MHz. The effect of the
spin-averaged terms is significant: Loss rates for CaF+CaF as a function of
magnetic field for $\Omega=100$~MHz are shown in Fig.~\ref{fig:CaF_highB}. For
high magnetic fields, the microwave-induced loss approaches a constant that is
significantly higher than the spin-free loss rate. This figure also includes
the loss rates obtained by averaging over a single spin-stretched state; they
agree quantitatively with the high-field results from full calculations
including all fine and hyperfine states. This again illustrates that the
suppression of fine and hyperfine interactions results from uncoupling of the
spin and rotational degrees of freedom by quantizing the spins along the
external magnetic field axis.

\end{document}